\documentclass[fleqn,12pt]{article}
\usepackage{epsfig}
\newcommand{\be}{\begin{equation}}
\newcommand{\ee}{\end{equation}}

\newcommand{\eq}[1]{eq.~(\ref{#1})}

\newcommand{\pbar}{\bar p}
\def\signn{\sigma_{\rm nn}}
\def\siggp{\sigma_{\gamma\rm p}}
\def\siggg{\sigma_{\gamma\gamma}}
\def\rhonn{\rho_{\rm nn}}
\def\rhogp{\rho_{\gamma\rm p}}
\def\rhogg{\rho_{\gamma\gamma}}
\def\Bnn{B_{\rm nn}}
\def\Bgp{B_{\gamma\rm p}}
\def\Bgg{B_{\gamma\gamma}}
\begin{document}    
\renewcommand\thepage{\ }
%
%
\begin{titlepage} 
%
\newcommand\reportnumber{1001} 
\newcommand\mydate{June 7, 2002} 
\newlength{\nulogo} 
\settowidth{\nulogo}{\small\sf{N.U.H.E.P. Report No. \reportnumber}}
\title{\hfill\fbox{{\parbox{\nulogo}{\small\sf{Northwestern University: \\
N.U.H.E.P. Report No. \reportnumber\\ \mydate
}}}}\vspace{1in} \\
{Forward Elastic Scattering of Light on Light, $\gamma+\gamma\rightarrow\gamma+\gamma$}}
 
\author{
Martin~M.~Block
\thanks{Work partially supported by Department of Energy contract
DA-AC02-76-Er02289 Task D.}\vspace{-5pt}   \\
{\small\em Department of Physics and Astronomy,} \vspace{-5pt} \\ 
{\small\em Northwestern University, Evanston, IL 60208}\\
\vspace{-5pt}
\  \\
Giulia~Pancheri
\vspace{-5pt} \\
{\small\em INFN-Laboratori Nazionali di Frascati,}\vspace{-5pt}  \\
{\small\em I00044 Frascati, Italy}\\
\vspace{-5pt}\\
}    
\vspace{.5in}
\date{} 
\maketitle
\begin{abstract} 
\noindent The forward elastic scattering of light on light, {\em i.e.,} the reaction $\gamma+\gamma\rightarrow\gamma+\gamma$ in the forward direction, is analyzed utilizing real analytic amplitudes. We calculate $\rho_{\gamma \gamma}$, the ratio of the real to the imaginary portion of the forward scattering amplitude, by fitting the total $\gamma \gamma$ cross section data in the high energy region $5\ {\rm GeV}\le \sqrt s\le 130 $ GeV, assuming a cross section that rises asymptotically as $\ln^2 s$. We then compare $\rho_{\gamma\gamma}$ to $\rho_{nn}$, the ratio of the {\em even} portions of the $pp$ and $\pbar p$ forward scattering amplitudes, as well as to $\rho_{\gamma p}$\cite{rhogp}, the $\rho$ value for Compton scattering.  Within errors, we find that the three $\rho$-values in the c.m.s. energy region $5 \ {\rm GeV}\le\sqrt s\le 130$ GeV are the same, as predicted by a factorization theorem of Block and Kadailov\cite{bk}.
\end{abstract}  
\end{titlepage} 
%
\pagenumbering{arabic}
\renewcommand{\thepage}{-- \arabic{page}\ --}  
%
The purpose of this note is to analyze light on light elastic scattering at high energies, {\em i.e.,} $\gamma+\gamma\rightarrow\gamma+\gamma$, when the scattered $\gamma$ is in the forward direction, in order to extract  $\rho_{\gamma \gamma}$, the ratio of the real to the imaginary portion of the forward scattering amplitude, by using real analytic amplitudes. To the best of our knowledge, no one has either measured or calculated $\rho_{\gamma \gamma}$. 

Block and Kadailov\cite{bk} have shown that $\rho_{\rm nn}=\rho_{\gamma p}=\rho_{\gamma \gamma}$ if one uses eikonals for $\gamma \gamma$, $\gamma p$ and  the even portion of nucleon-nucleon scattering that have {\em equal} opacities, {\em i.e.,} eikonals that have the same value at impact parameter $b=0$ (for details of the assumptions made, see ref. \cite{bk}). This is the equivalent of the more physical statement that
\be
\left(
\frac{\sigma_{\rm elastic}(s)}{\sigma_{\rm tot}(s)}
\right)_{\gamma \gamma}=
\left(\frac{\sigma_{\rm elastic}(s)}{\sigma_{\rm tot}(s)}
\right)_{\gamma p}
=\left(\frac{\sigma_{\rm elastic}(s)}{\sigma_{\rm tot}(s)}\right)_{nn},\quad {\rm for\  all\ }s.\label{sigratios}
\ee
Block and Kaidalov\cite{bk} have proved three  factorization theorems:
\begin{equation}
 \frac{\signn(s)}{\siggp(s)}=\frac{\siggp(s)}{\siggg(s)},\label{eq:sig}
\end{equation}
 where the $\sigma$'s are the total cross sections for nucleon-nucleon, $\gamma$p and $\gamma\gamma$ scattering,
\begin{equation}
\frac{\Bnn(s)}{\Bgp(s)}=\frac{\Bgp(s)}{\Bgg(s)},\label{eq:B}
\end{equation}
 where the $B$'s are the nuclear slope parameters for elastic scattering, and 
\begin{equation} \rhonn(s)=\rhogp(s)=\rhogg(s),\label{eq:rho}
\end{equation}
 where the $\rho$'s are the ratio of the real to imaginary portions of the forward scattering amplitudes,  
with the first two  factorization theorems each having their own proportionality constant.  These theorems are exact, for {\em all } $s$  (where $\sqrt s$ is the c.m.s. energy), and survive exponentiation of the eikonal (see ref. \cite{bk}). We emphasize that the equality of the three $\rho$ values, the theorem of \eq{eq:rho},  is valid {\em independently} of the model which takes one from $nn$ to $\gamma p$ to $\gamma\gamma$ reactions, as long as the respective eikonals have equal opacities, {\em i.e.,} \eq{sigratios} holds.  Block\cite{rhogp} has already shown that $\rhonn(s)=\rho_{\gamma p}(s)$. 

The purpose of this communication is to demonstrate using available high energy experimental data,  that within errors, $\rhonn(s)=\rho_{\gamma\gamma}(s)$, completing the validation of the last factorization theorem given in \eq{eq:rho}.   However, no direct measurements are available for $\rho_{\gamma\gamma}(s)$.  We will find $\rho_{\gamma\gamma}(s)$ utilizing an analysis involving real analytic amplitudes, a technique first proposed by Bourrely and Fischer\cite{bourrely}. This work follows the procedures and conventions used by Block and Cahn\cite{bc}. The variable $s$ is the square of the c.m. system energy, whereas $\nu$ is the laboratory system momentum. In terms of the {\em even}  laboratory scattering amplitude $f_+$, where $f_+(\nu)=f_+(-\nu)$, the
total unpolarized $\gamma\gamma$ cross section $\sigma_{\gamma\gamma}$ is given by
\begin{eqnarray}
\sigma_{\gamma\gamma}&=&\frac{4\pi}{\nu}{\rm Im}f_+(\theta=0),\label{optical}
\end{eqnarray}
where $\theta$ is the laboratory scattering angle.
We further assume that our amplitudes are real analytic functions with a simple cut structure\cite{bc}. We use an even amplitude for $\gamma \gamma$ reactions in the high energy region, far above any cuts,  (see ref.\cite{bc}, p. 587, eq. (5.5a), with $a=0$), where the even amplitude simplifies considerably and is given by
\begin{equation}
f_+(s)=i\frac{\nu}{4\pi}\left\{A+\beta[\ln (s/s_0) -i\pi/2]^2+cs^{\mu-1}e^{i\pi(1-\mu)/2}\right\},\label{evenamplitude_gp}
\end{equation}
where $A$, $\beta$, $c$, $s_0$ and $\mu$ are real constants. We are ignoring any real subtraction constants. In \eq{evenamplitude_gp}, we have assumed that the $\gamma\gamma$  cross section rises asymptotically as $\ln^2 s$. The real and imaginary parts of \eq{evenamplitude_gp} are given by
\begin{eqnarray}
{\rm Re}\frac{4\pi}{\nu}f_+(s) &=&\beta\,\pi \ln s/s_0-c\,\cos(\pi\mu/2)s^{\mu-1}\label{real}\\ 
{\rm Im}\frac{4\pi}{\nu}f_+(s) &=&A+\beta\left[\ln^2 s/s_0-\frac{\pi^2}{4}\right]+c\,\sin(\pi\mu/2)s^{\mu-1}. \label{imaginary}
\end{eqnarray}
 Using equations ({\ref{optical}), (\ref{real}) and  (\ref{imaginary}), the total cross section for high energy $\gamma\gamma$ scattering is given by
\be
\sigma_{\rm tot}(s)= A+\beta\left[\ln^2 s/s_0-\frac{\pi^2}{4}\right]+c\,\sin(\pi\mu/2)s^{\mu-1} , \label{sigmatot}
\ee 
and $\rho$, the ratio of the real to the imaginary portion of the forward scattering amplitude, is given by
\be
\rho(s)=\frac{\beta\,\pi\ln s/s_0-c\,\cos(\pi\mu/2)s^{\mu-1}}{\sigma_{\rm tot}}.\label{rho}
\ee  
If we assume that the  term in $c$ is a Regge descending term, then $\mu=1/2$.  
 
Total cross sections for $\gamma \gamma$ scattering are now available for c.m.s. energies up to $\approx $130 GeV\cite{exp}. We have made a $\chi^2$ fit of \eq{sigmatot} to the experimental $\sigma_{\rm tot}(\gamma \gamma)$ data in the c.m.s. energy interval $5 {\ \rm GeV}\le \sqrt s\le 130$ GeV.  We find a reasonable representation of the data using \eq{sigmatot}, with a $\chi^2$ per degree of freedom of 0.066 for 9 degrees of freedom, with the coefficients: 

$A= 377\pm 16$ nb,  $\beta=11.1\pm 5.3$ nb, $s_0=266.7\pm 157$ GeV$^2$, 

\noindent using the fixed values, $c=220$ nbGeV and $\mu=0.5$. The cross sections of \eq{sigmatot} are in nb, with $s$ in GeV$^2$. This fit, plotted as a function of c.m.s. energy, gives the dashed cross section curve $\sigma_{\rm tot}(\gamma \gamma)$ in  Fig. \ref{fig:siggammap}, as well as the dashed  $\rho_{\gamma \gamma}$ curve in Fig. \ref{fig:rhogammagamma}, using \eq{sigmatot} and \eq{rho}, respectively. Several comments about both the experimental data and the fit are in order. First, the two  independent  cross section measurements of L3 and OPAL have been utilized, which are  separately normalized using Monte Carlos. Second, the expected $\chi^2$/d.f. is 1, whereas the value obtained in our fit is only 0.066---a very unlikely result if the experimental points are uncorrelated and have their statistical errors allocated correctly.  Clearly, more precise data is required to settle this issue, as well as reducing the (rather large) errors of the fitted parameters. Third, it should be emphasized that the $\rho$ value (see \eq{rho}) is independent of the  absolute normalization of the total cross section $\sigma_{\rm tot}$, depending  only on its shape. Clearly,  relative normalization errors between the OPAL and L3 experiments play a critical role in shape dependence, whereas  overall normalization does not affect the result for $\rho$.   

In order to demonstrate the sensitivity of $\sigma_{\rm tot}(\gamma \gamma)$ and $\rho_{\gamma \gamma}$ to the parameters of the fit, we have also plotted in Fig. \ref{fig:siggammap} the dotted curve (a variation of the parameters of $s_0$ and $\beta$ within their errors), where we have set $s_0=189$ GeV$^2$ and $\beta=5.9$.  This curve has as its $\rho_{\gamma \gamma}$ analog the dotted curve of Fig. \ref{fig:rhogammagamma}.  
\begin{figure}[p] 
\begin{center}
\mbox{\epsfig{file=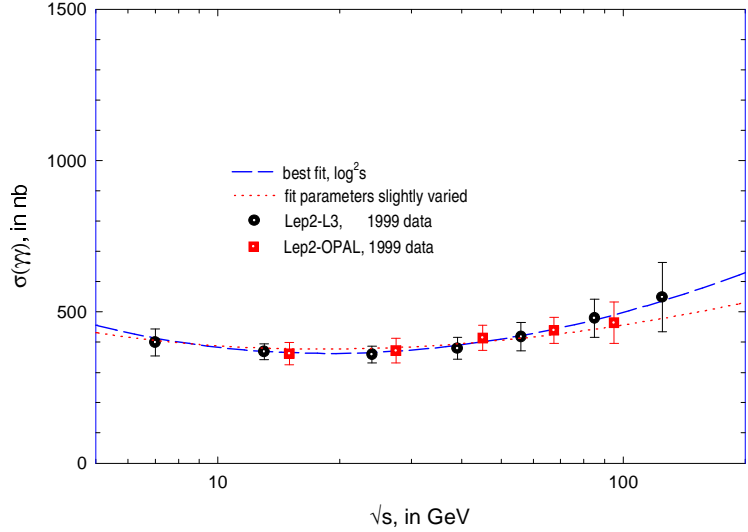,width=4.4in,%
bbllx=85pt,bblly=380pt,bburx=520pt,bbury=665pt,clip=}}
\end{center}
\caption[] {\footnotesize The dashed curve is $\sigma_{\rm tot}(\gamma \gamma)$, the predicted total $\gamma \gamma$ cross section (in nb) from \eq{sigmatot}, using the central value parameters of a $\chi^2$ fit, $A= 377$ nb,  $\beta=11.1$ nb, $s_0=266.7$ GeV$^2$, with $\mu=0.5$ and $c=220$ nbGeV, compared to the existing high energy experimental data in the c.m.s. energy interval $5 {\ \rm GeV}\le \sqrt s\le 130$ GeV.  The dotted curve varies the parameters slightly, with $s_0\rightarrow 189$ GeV$^2$ and $\beta\rightarrow 5.9$ nb (values within their errors). The corresponding $\rho_{\gamma \gamma}$ curves are shown in Fig. \ref{fig:rhogammagamma}.}
\label{fig:siggammap}
\end{figure}
%
\begin{figure}[p] 
\begin{center}
\mbox{\epsfig{file=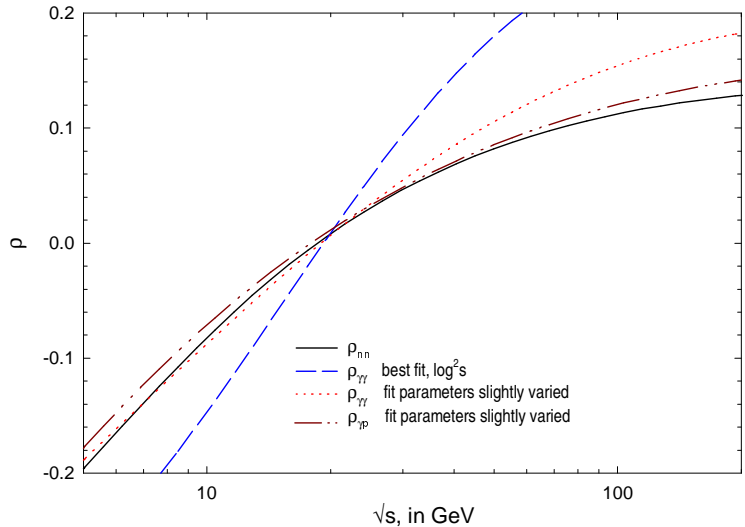,width=4.4in,%
bbllx=85pt,bblly=380pt,bburx=520pt,bbury=665pt,clip=}}
\end{center}
\protect\caption[] {\footnotesize The solid curve is $\rho_{\rm nn}$, the predicted ratio of the 
real to imaginary part of the forward scattering amplitude for the
`elastic reactions , $\gamma +\gamma\rightarrow V_i + V_j$ scattering
amplitude, where $V_{i,j}$ is $\rho$, $\omega$ or $\phi$ (using  factorization).
The dashed curve is $\rho_{\gamma \gamma}$, the ratio of the real to imaginary
part of the forward scattering amplitude for elastic 
scattering, $\gamma +\gamma\rightarrow\gamma + \gamma$, 
found from \eq{rho} using real analytic amplitudes that asymptotically go as $\ln^2 s$, with best fit parameters $A= 377$ nb,  $\beta=11.1$ nb, $s_0=266.7$ GeV$^2$, with $c=220$ nbGeV and $\mu=0.5$.  The dotted line is the $\rho_{\gamma \gamma}$ curve  where the parameters of the fit have been slightly varied within their errors, with  $s_0\rightarrow 189$ GeV$^2$  and $\beta\rightarrow 5.9$ nb.  The corresponding two curves for  $\sigma_{\rm tot}(\gamma \gamma)$ are shown in Fig. \ref{fig:siggammap}. The dashed dot dot curve is $\rho_{\gamma p}$, taken from ref. \cite{rhogp} and is shown for comparison with $\rho_{\gamma\gamma}$ and $\rho_{\rm nn}$.
\label{fig:rhogammagamma}}
\end{figure}
Using \eq{rho}, Fig. \ref{fig:rhogammagamma} shows our result for $\rho_{\gamma \gamma}$  compared to $\rho_{nn}$, the $\rho$-value for the even portion of nucleon-nucleon scattering found in ref. \cite{blockcr}, as a function of the c.m.s. energy $\sqrt s$, in GeV. The solid curve is $\rho_{nn}$; the dashed line  is the $\rho_{\gamma \gamma}$ curve which corresponds to the central values $A= 377$ nb, $\beta=11.1$ nb, $s_0=266.7$ GeV$^2$, with $c=220$ nbGeV and $\mu=0.5$; the dotted line is the $\rho_{\gamma \gamma}$ curve which uses the slightly varied parameters $s_0=189$ GeV$^2$ and $\beta=5.9$ nb. Also shown as the dashed dot dot curve in Fig. \ref{fig:rhogammagamma} is $\rho_{\gamma p}(s)$, taken from a recent  analysis of Compton scattering\cite{rhogp}.   
The agreement between the slightly modified $\rho_{\gamma \gamma}(s)$ and $\rho_{nn}(s)$ over the energy interval $5\ {\rm GeV}\le \sqrt s\le 130$ GeV, as well as with as with $\rho_{\gamma p}(s)$, lends experimental support---in a model independent way---for the three factorization theorems of Block and Kaidalov\cite{bk,bhp}, $\signn(s)/\siggp(s)=\siggp(s)/\siggg(s)$, $\Bnn(s)/\Bgp(s)=\Bgp(s)/\Bgg(s)$, and $\rhonn(s)=\rhogp(s)=\rhogg(s)$. Clearly, these conclusions would be greatly strengthened by precision cross section measurements of  both $\gamma p$ and $\gamma \gamma$ reactions at high energies. 
\end{document}